# Flash-Radiomics: A Scalable Hybrid CPU–CUDA Engine for Standardized Scalar Radiomics and Accelerated Spatial Mapping

Shanli Ding[1,2], Yiyi Hu[3], Ziyu Fu[1,2], Chia-Hsin Lin[1,4], Ruihan Luo[5], Jaehee Chun[2], Xinyue Zhang[2,6], Osama Mawlawi[2]

1. Graduate School of Biomedical Sciences, The University of Texas MD Anderson Cancer Center UTHealth Houston, Houston, TX 77030, USA
2. Department of Imaging Physics, The University of Texas MD Anderson Cancer Center, Houston, TX 77030, USA
3. Division of Nuclear Medicine and Molecular Imaging, Geneva University Hospital, Geneva, Switzerland
4. Department of System Biology, The University of Texas MD Anderson Cancer Center, Houston, TX 77030, USA
5. Gastroenterology Research Center (GRC)/Department of Internal Medicine, The University of Texas Health Science Center at Houston, Houston, Texas, USA
6. Department of Computer Science, Rice University, Houston, TX 77005, USA

**Background and Objectives:** Spatial mapping retains the spatial distribution of radiomic features, but computational cost and fragmented software limit its use. We developed Flash-Radiomics with scalar extraction and spatial mapping, a central processing unit (CPU) backend, a hybrid Compute Unified Device Architecture (CUDA) backend, consistent feature names, and Hierarchical Data Format version 5 (HDF5) storage.

**Methods:** We evaluated Image Biomarker Standardisation Initiative (IBSI) compliance, CPU-CUDA concordance, and end-to-end processing time. Compliance testing included 825 chapter 1 (IBSI-1) tests covering 165 high-consensus features and 323 chapter 2 (IBSI-2) tests with numerical references. Concordance testing included 1,148 scalar pairs and 93 spatial-map pairs. End-to-end processing time was measured five times per input volume of interest (VOI) size. Comparisons included the Medical Image Radiomics Processor (MIRP) and PyRadiomics for 102 shared scalar features and PyRadiomics for 93 shared spatial maps.

**Results:** Both backends passed all 1,148 IBSI tests, and all paired results were concordant. At the largest scalar input, CPU required 76.343 s and hybrid CUDA 81.915 s; CPU was 4.7 times faster than MIRP and 190.7 times faster than PyRadiomics. At the largest spatial input completed by both backends, hybrid CUDA reduced processing time by 68.6% relative to CPU (79.280 versus 252.791 s). At PyRadiomics' largest completed spatial input, hybrid CUDA was 84.9 times faster.

**Conclusions:** Flash-Radiomics unified standardized scalar extraction, spatial mapping, concordant CPU-CUDA results, and HDF5 storage. CPU processing time was similar or shorter for scalar extraction, whereas hybrid CUDA was faster for spatial mapping under the tested conditions.

## Introduction

Medical imaging plays a central role in clinical assessment, yet much of its interpretation remains qualitative. Radiomics complements this practice by converting images into quantitative descriptors of tissue phenotype and intralesional heterogeneity [1–3]. These structured measurements can then be incorporated into statistical and computational analyses of medical images.

Radiomics can be performed in two extraction modes: scalar extraction and spatial mapping. Scalar extraction returns one value per feature for a volume of interest (VOI), providing a compact summary of the complete volume. Spatial mapping calculates features within local image neighborhoods and returns maps aligned with the source image. Because each map value describes a local image region, spatial mapping shows where patterns differ within the VOI and can support analyses such as radiomic habitat characterization [4]. The two extraction modes therefore provide complementary information. However, spatial mapping requires many more calculations and produces much larger results than scalar extraction.

Existing radiomics tools differ in their support for scalar extraction, spatial mapping, and graphics processing unit (GPU) acceleration. For example, PyRadiomics provides 102 scalar features and 93 three-dimensional (3D) spatial maps [5], whereas the Medical Image Radiomics Processor (MIRP) provides 165 scalar features with no spatial maps [6]. The Computational Environment for Radiological Research (CERR) provides 86 scalar and 35 spatial features in MATLAB and Octave [7]. Local Image Feature Extraction (LIFEx) documents 111 scalar features and spatial texture maps, although the number of maps is not stated [8,9]. Tools based on the Compute Unified Device Architecture (CUDA) have narrower feature coverage: PyRadiomics-CUDA accelerates 23 morphology features within the 102-feature PyRadiomics inventory [10], cuRadiomics implements 18 first-order and 23 gray-level co-occurrence matrix (GLCM) scalar features [11,12], and HaraliCU generates 16 two-dimensional (2D) spatial maps [13–15]. The desktop application Imaging Biomarker Explorer (IBEX) documents 80 scalar features [16]. Table 1 summarizes feature coverage, central processing unit (CPU) and GPU support, and two forms of software access. An application programming interface (API) allows another program to request features and receive results directly, whereas an automated batch interface processes multiple image–mask inputs without repeated manual interaction.

### Table 1. Feature coverage, GPU scope, interfaces, and Image Biomarker Standardisation Initiative (IBSI) compliance

Table note: [1]A check mark indicates that the tool passed all 825 IBSI chapter 1 (IBSI-1) tests derived from 165 high-consensus features and all 323 evaluated IBSI chapter 2 (IBSI-2) tests. A dash means that a complete pass was not demonstrated in this study; it does not indicate noncompliance. Flash-Radiomics provides 173 scalar features, including the 165-feature IBSI-1 set. [2]The IBEX and cuRadiomics totals were not matched one-to-one with this set. [3]LIFEx does not report the number of its spatial maps. [4]GPU-enabled feature counts describe the features or maps available through a CUDA implementation. The Flash-Radiomics counts apply to its hybrid CUDA backend; the PyRadiomics-CUDA count is its accelerated morphology subset. A dash indicates no documented GPU feature execution. Spatial counts are not IBSI compliance counts. Version numbers are shown only for the comparator releases executed in this study; the other comparator rows summarize the cited public implementations. The table does not rank correctness, clinical value, or speed.

| Tool | Runtime | Scalar Features | Spatial Features | GPU-enabled features[4] | IBSI-compliant[1] | Programmatic API | Automated batch interface |
|---|---|---|---|---|---|---|---|
| **Flash-Radiomics** | Python/C/CUDA | 165 | 101 | ✓ | ✓ | ✓ | ✓ |
| PyRadiomics v3.0.1 [5] | Python | 102 | 93 | — | — | ✓ | ✓ |
| PyRadiomics-CUDA v1.0.4 [10] | Python/CUDA | 102 | — | ✓ (23 morphology) | — | ✓ | ✓ |
| MIRP v2.5.0 [6] | Python | 165 | — | — | ✓ | ✓ | ✓ |
| CERR [7] | MATLAB/Octave | 86 | 35 | — | — | ✓ | ✓ |
| IBEX v1.0β [16] | MATLAB/C++ desktop | 80[2] | — | — | — | — | — |
| LIFEx v25.06.1 [8,9] | Desktop executable | 111 | Not stated[3] | — | — | — | ✓ |
| cuRadiomics [11,12] | Python/CUDA | 41[2] | — | ✓ (18 first-order; 23 GLCM) | — | ✓ | — |
| HaraliCU [13-15] | C++/CUDA executable | — | 16 (2D) | ✓ | — | — | ✓ |

As shown in Table 1, the relevant capabilities remain divided among the other reviewed tools. Software with broad scalar or 3D spatial coverage does not provide GPU execution, whereas GPU-enabled tools support limited feature groups or only 2D maps. No other reviewed tool combines broad scalar extraction, 3D spatial mapping, CPU and CUDA execution, a programmatic API, and automated batch processing in one workflow. This fragmentation makes spatial radiomics harder to apply to large images and broad feature sets.

Reliable radiomics also requires consistent numerical definitions. Feature values can change with image preparation, VOI handling, intensity discretization, feature equations, and numerical implementation. The Image Biomarker Standardisation Initiative (IBSI) provides common terminology, definitions, and reference values for checking software implementations [17,18]. However, software can use different feature names, supported feature groups, and calculation conventions. Matching names alone therefore does not guarantee matching values [19,20].

In an effort to overcome these drawbacks, we developed Flash-Radiomics to provide scalar extraction and spatial mapping with consistent feature names through native CPU and hybrid CUDA backends. Both backends use the same processing settings and feature definitions. A single compressed Hierarchical Data Format version 5 (HDF5) file can store the source image, mask, geometry, extraction settings, feature names, and scalar values or spatial maps together. This design keeps the information for an extraction in one organized file for downstream analysis.

We evaluated Flash-Radiomics in three sequential analyses. First, we assessed both backends against IBSI numerical references and nomenclature using 825 IBSI-1 tests derived from 165 high-consensus features and 323 IBSI-2 tests. We compared the resulting test outcomes with those available from MIRP and PyRadiomics. Second, we evaluated CPU-CUDA concordance for scalar values and spatial maps. Third, we compared end-to-end processing time for 102 scalar features shared by Flash-Radiomics, MIRP, and PyRadiomics and for 93 spatial maps shared by Flash-Radiomics and PyRadiomics.

## Materials and methods

## Software design and extraction modes

Flash-Radiomics provides two extraction modes. Throughout this study, extraction mode refers only to scalar extraction or spatial mapping, and CPU and CUDA refer to execution backends. Scalar extraction returns one value for each requested feature and VOI. Spatial mapping returns a 3D map aligned with the source-image geometry, and each map location describes an image neighborhood.

Flash-Radiomics accepts a medical image, segmentation mask, extraction settings, and selected backend. The settings define the image types, feature families, intensity discretization, VOI handling, and extraction mode. Python and SimpleITK, a simplified interface to the Insight Segmentation and Registration Toolkit, manage image input, geometry, preprocessing, and image interpretation. After validating the VOI and preparing contiguous arrays, Flash-Radiomics sends the requested calculations to the CPU or CUDA backend. Its Python API returns the results as a feature dictionary. Because some features are available in only one extraction mode, a shared family name does not mean that the scalar and spatial feature lists are identical. Figure 1 summarizes these steps.

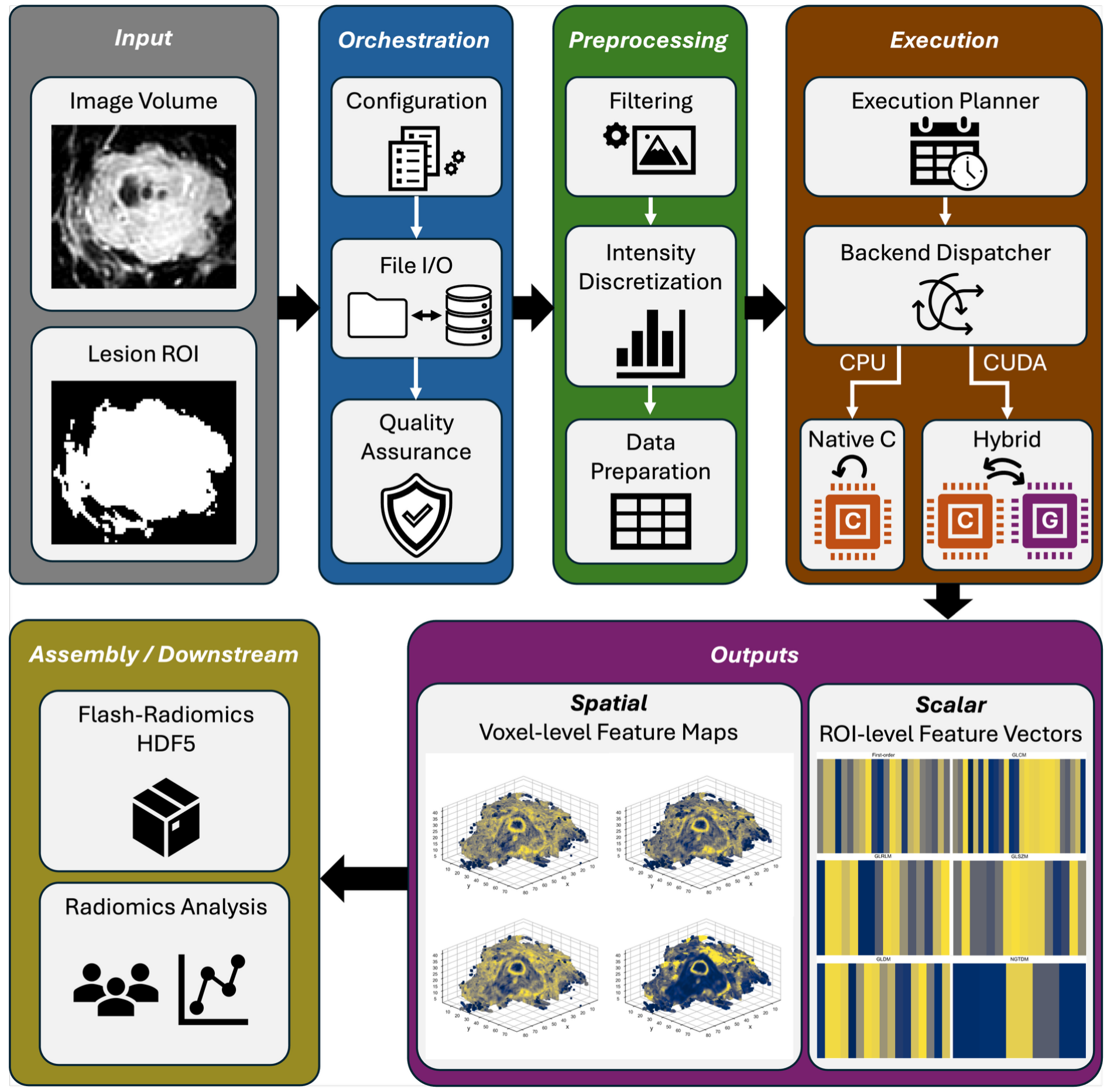

**Figure 1. Flash-Radiomics workflow. Python and SimpleITK prepare the image, VOI mask, geometry, and extraction settings. The selected backend calculates the requested scalar values or spatial maps, which can be saved with the input information in one HDF5 file.**

## CPU backend implementation

The CPU implementation separates image preparation from feature calculation. Python and SimpleITK prepare the image and mask arrays and perform shared preprocessing once. Native C functions then calculate only the requested features. These functions construct histograms and texture matrices directly, reuse allocated memory, track occupied intensity bins, and clear only the state that must be reset. This design reduces repeated preparation and memory allocation.

## Hybrid CUDA backend implementation

The hybrid CUDA backend uses GPU implementations for supported calculations while retaining the same feature definitions as the CPU backend. Feature families offer different amounts of parallel work, and some operations still require CPU execution. Image preparation, data transfer, GPU function calls, and calculations that remain on the CPU all contribute to end-to-end processing time. Selecting the CUDA backend therefore does not mean that every calculation runs on the GPU.

For spatial mapping, one function invocation from Python starts the CUDA calculation, after which the enabled feature classes are processed in sequence. The current implementation does not share GPU uploads among feature classes and does not run feature classes concurrently in separate CUDA streams. Preprocessing, intensity discretization, connected-component operations, and exact fallback calculations remain on the CPU when required. CPU and CUDA calculations can therefore be used within the same extraction while preserving the requested feature definitions.

## Structured HDF5 storage

Flash-Radiomics can save each extraction in one HDF5 file [21] without changing the feature dictionary returned by the API. For either extraction mode, the file stores the results together with the image and mask arrays, voxel geometry, extraction settings, and information identifying the input. Scalar results are stored as ordered feature names and 64-bit floating-point values. Spatial results are stored as named maps with their source geometry. The user can select Lempel-Ziv Fast or gzip compression in the extraction settings.

PyRadiomics documents tabular scalar output and separate Nearly Raw Raster Data (NRRD) files for spatial maps [22], whereas MIRP returns scalar features as a pandas DataFrame [23]. These formats remain suitable for individual-map exchange and tabular analysis. Flash-Radiomics instead uses HDF5 to centralize the image, mask, geometry, settings, provenance, and scalar or spatial results in one structured file. This comparison concerns output organization; speed, storage size, and reproducibility were not compared among formats.

## IBSI compliance tests

The first numerical experiment evaluated IBSI compliance by comparing scalar results with IBSI reference values [17,18]. IBSI chapter 1 (IBSI-1) covers conventional radiomic features, whereas IBSI chapter 2 (IBSI-2) provides reference values for statistical features calculated after standardized image filtering.

The IBSI compliance analysis was limited to scalar outputs. The evaluated IBSI-1 and IBSI-2 benchmarks provide scalar reference values but no standardized spatial radiomic feature maps for map-to-map compliance testing [17,18]. In IBSI-2, the scalar values are statistics derived from standardized filter-response images [18]; these response images should not be confused with spatial radiomic feature maps. Spatial maps were therefore excluded from this analysis and evaluated separately by CPU-CUDA concordance.

The IBSI-1 test included five configurations and 165 high-consensus features. A configuration denotes an IBSI-defined test condition that specifies the reference image and VOI mask, processing dimensionality, interpolation and resampling, resegmentation, intensity discretization, and texture-feature aggregation. One reference value was available for each feature under each configuration, giving 825 tests. When an IBSI reference quantity did not have a one-to-one output in the Flash-Radiomics feature dictionary, the test program calculated it from direct shape results and image or mask information for that configuration. These quantities included volume, surface area, sphericity, mean intensity, centers of mass, and bounding boxes. Each result was recorded as direct or derived.

The IBSI-2 test included 22 filter configurations and 18 statistical features, giving 396 possible tests. We included the 323 tests for which IBSI provides a numerical reference value. The other 73 tests did not have a numerical reference value and were excluded before the pass rate was calculated. Subsequent references to the evaluated IBSI-2 tests are attributed to these 323 comparisons.

For each test, we used the tolerance specified by IBSI and followed the comparison procedure supplied with the corresponding benchmark. The IBSI-1 submission template evaluates the numerical difference at the reporting precision of the reference value before applying its tolerance [24]. For example, a calculated value of 0.0454545455 matches a reference value reported as 0.0455 when both are compared at the four displayed decimal places. For IBSI-2, the absolute difference was compared directly with the published tolerance because every included reference had a positive numerical tolerance [18].

Let $d_{\mathrm{IBSI}}$ denote the difference obtained with the relevant IBSI comparison procedure. A result passed when

$$d_{\mathrm{IBSI}} \leq T_{\mathrm{IBSI}},$$

where $T_{\mathrm{IBSI}}$ is the tolerance specified by IBSI.

The same reference values and comparison procedure were applied to both Flash-Radiomics backends and the other applicable tools. Flash-Radiomics CPU, Flash-Radiomics CUDA, and MIRP were tested against all 825 IBSI-1 references. For PyRadiomics, each feature name and configuration setting was matched to an IBSI target with the same mathematical definition. This procedure identified 102 comparable features in each of the five configurations, giving 510 numerical comparisons. The other 315 IBSI-1 tests were classified as unsupported, not failed.

Flash-Radiomics CPU, Flash-Radiomics CUDA, and MIRP were also tested against all 323 evaluated IBSI-2 references. PyRadiomics was tested only when its image type and statistical-feature definition matched the IBSI definition. It supported 15 statistical features in each of the two configurations without image filtering, giving 30 comparisons. The other 293 IBSI-2 tests required unavailable features or different filters and were classified as unsupported.

Test results of passed, failed, and unsupported tests were reported as proportions of the complete reference set so that numerical test outcomes and feature coverage remained visible separately. The primary comparison used the values returned by each tool and the IBSI comparison procedure described above; no feature value was otherwise transformed. In a secondary analysis, we subtracted three from the matched PyRadiomics kurtosis values and repeated the comparison. This check tested whether differences arose from conventional versus excess kurtosis and did not change the primary classifications. Other failures were grouped by IBSI chapter, configuration, and feature family and compared with documented PyRadiomics feature and discretization definitions [25].

## CPU-CUDA concordance tests

The second numerical experiment evaluated CPU-CUDA concordance for scalar values and spatial maps. Passing the same external references does not ensure identical results between backends because CPU and CUDA calculations can evaluate floating-point operations in different orders. The scalar test used the same 825 IBSI-1 and 323 evaluated IBSI-2 tests, giving 1,148 CPU-CUDA pairs. For a CPU result $x_{\text{CPU}}$ and CUDA result $x_{\text{CUDA}}$, the absolute difference was

$$d_{\text{abs}} = |x_{\text{CUDA}} - x_{\text{CPU}}|$$

and the relative difference was

$$d_{\text{rel}} = \frac{|x_{\text{CUDA}} - x_{\text{CPU}}|}{\max(|x_{\text{CPU}}|, |x_{\text{CUDA}}|, 1)}$$

A pair was classified as exact when $d_{\text{abs}} = 0$. A non-exact pair was considered concordant when at least one of the following criteria was met:

$$d_{\text{abs}} \leq 1 \times 10^{-6} \quad \text{or} \quad d_{\text{rel}} \leq 1 \times 10^{-11}$$

These study-defined limits cover complementary magnitude regimes. The OR rule is equivalent to

$$d_{\text{abs}} \leq \max[1 \times 10^{-6}, 1 \times 10^{-11} \max(|x_{\text{CPU}}|, 1)]$$

The absolute tolerance therefore governs small-magnitude values, for which conventional relative differences can be unstable. The relative tolerance prevents large-magnitude features from being rejected because of negligible proportional differences. Requiring both limits would instead apply the smaller allowance, negating the absolute floor near zero and the relative scaling at large magnitudes. The CPU result served as the reference, and the denominator floor of 1 stabilized the relative measure near zero.

The spatial concordance test used the 93-map feature set and an expert lesion mask with its corresponding image from a public breast cancer magnetic resonance imaging (MRI) dataset [26]. The source image contained 448 × 448 × 208 voxels, with a voxel spacing of 0.78125 × 0.78125 × 1.0 mm and a physical extent of 350.00 × 350.00 × 208.00 mm. We cropped the image and mask to 61 × 55 × 39 voxels, corresponding to 47.66 × 42.97 × 39.00 mm. The cropped mask contained 50,613 VOI voxels, corresponding to 30.9 mL. The same crop was used to construct the computational-performance inputs.

Each CPU-CUDA spatial map pair first had to have identical geometry and the same pattern of finite and nonfinite values. At every finite location, we calculated $d_{\text{abs}}$ and $d_{\text{rel}}$ using the definitions above. We then applied the same OR criterion and the same $1 \times 10^{-6}$ absolute and $1 \times 10^{-11}$ relative limits used for scalar results. A spatial map was considered concordant only when every finite value pair met at least one limit. This test measured concordance between the two backends for the evaluated image and mask. It did not establish accuracy against an external standardized spatial map or treat map values as independent samples.

## End-to-end processing-time evaluation

The computational-performance comparisons required each tool to perform the same preprocessing and produce the same type and number of results through an automated procedure. For this test, we considered PyRadiomics [5], PyRadiomics-CUDA [10], MIRP [6], CERR [7], the Imaging Biomarker Explorer (IBEX) [16], LIFEx [8,9], cuRadiomics [11,12], and HaraliCU [13–15] (Table 1).

PyRadiomics was included in the scalar and spatial comparisons because it supports programmatic scalar extraction and 3D spatial maps [5]. MIRP was included in the scalar comparison because it provides the full standardized IBSI feature coverage [6], but it does not generate spatial maps. The remaining tools were excluded because they did not meet all requirements for these performance tests. In particular, the GPU-enabled tools did not provide the same scalar and spatial feature sets required for both comparisons [10–15]. Their exclusion from the performance analysis was not a judgment of correctness, software quality, or clinical value.

The scalar comparison used 102 features available in Flash-Radiomics, PyRadiomics, and MIRP. The spatial comparison used 93 maps available in Flash-Radiomics and PyRadiomics. Both sets contained statistics, gray-level co-occurrence matrix (GLCM), gray-level run-length matrix (GLRLM), gray-level size-zone matrix (GLSZM), neighboring gray-level dependence matrix (NGLDM), and neighboring gray-tone difference matrix (NGTDM) features. Intensity histogram, intensity volume histogram, and gray-level distance-zone matrix (GLDZM) features were excluded because PyRadiomics does not provide corresponding classes. Morphology was excluded because it has no spatial-map counterpart. Matching the feature count prevented a tool from appearing faster simply because it calculated fewer results. Table 2 lists the complete Flash-Radiomics counts and the shared sets used in the comparisons.

**Table 2. Flash-Radiomics feature counts and shared feature sets used in the numerical and end-to-end computational processing timing performance tests**

| Feature family | Default Scalar Count | Scalar Benchmark | Default Spatial Count | Spatial Benchmark | High-consensus IBSI |
|---|---|---|---|---|---|
| Morphology | 26 | 0 | — | 0 | 23 |
| Statistics | 22 | 22 | 21 | 18 | 18 |
| Intensity histogram | 23 | 0 | — | 0 | 23 |
| Intensity volume histogram | 6 | 0 | — | 0 | 6 |
| GLCM | 26 | 26 | 26 | 24 | 25 |
| GLRLM | 16 | 16 | 16 | 16 | 16 |
| GLSZM | 16 | 16 | 16 | 16 | 16 |
| GLDZM | 16 | 0 | — | 0 | 16 |
| NGLDM | 17 | 17 | 17 | 14 | 17 |
| NGTDM | 5 | 5 | 5 | 5 | 5 |
| **Total** | **173** | **102** | **101** | **93** | **165** |

Table note: A dash means that the family has no spatial implementation; it does not mean that the scalar implementation is absent. The IBSI-1 column contains 165 high-consensus features selected from the 169 standardized IBSI-1 features. IBSI-2 is not shown separately because it tested 18 statistical features under 22 filter configurations; 323 tests had numerical reference values. The performance columns contain currently supported results shared by the tools in each comparison and are smaller than the complete Flash-Radiomics lists of 173 scalar features and 101 spatial maps.

Abbreviations: GLCM, gray-level co-occurrence matrix; GLRLM, gray-level run-length matrix; GLSZM, gray-level size-zone matrix; GLDZM, gray-level distance-zone matrix; NGLDM, neighboring gray-level dependence matrix; NGTDM, neighboring gray-tone difference matrix; VOI, volume of interest.

The computational-performance tests measured end-to-end processing time at increasing VOI sizes and examined whether the faster backend differed between the two extraction modes. Scalar extraction used the shared 102-feature set, and spatial mapping used the shared 93-map set. Separate sets were used because scalar extraction returns one value per feature, whereas spatial mapping returns a dense 3D map for each feature. Every implementation calculated the applicable shared set.

We increased the input lesion size by repeating the cropped image and mask along all three spatial axes without interpolation. The repeated volumes retained the source voxel spacing, image values, and mask pattern. They were controlled computational inputs and did not represent anatomically enlarged lesions or independent clinical observations. Scalar extraction was evaluated at eight VOI sizes: 50,613; 101,226; 404,904; 607,356; 1,214,712; 2,277,585; 6,073,560; and 10,122,600 voxels. Spatial mapping was evaluated at six of these sizes: 50,613; 101,226; 404,904; 607,356; 1,214,712; and 10,122,600 voxels. The spatial comparisons used a 5,400-s time limit. Figure 2 illustrates how the inputs were constructed.

Each implementation was run five times at each input size. We used the median end-to-end processing time as the main value and the minimum and maximum to show variation among the five measurements. CPU and CUDA were tested with the same input and feature set at each size. End-to-end processing time included Python control, preprocessing, feature calculation, CPU-GPU data transfer, CPU fallback calculations, result assembly, and result storage. The analysis did not separate calculation time from file writing and therefore did not compare storage formats. Comparisons with another tool were interpreted only when preprocessing, feature definitions, result type, result count, and execution conditions were aligned.

To summarize how time changed with VOI size, we fitted

$$t = aN^b,$$

where $t$ is the median end-to-end processing time, $N$ is the number of VOI voxels, and $a$ and $b$ are fitted values. The fit described the tested inputs and was not used to predict clinical performance.

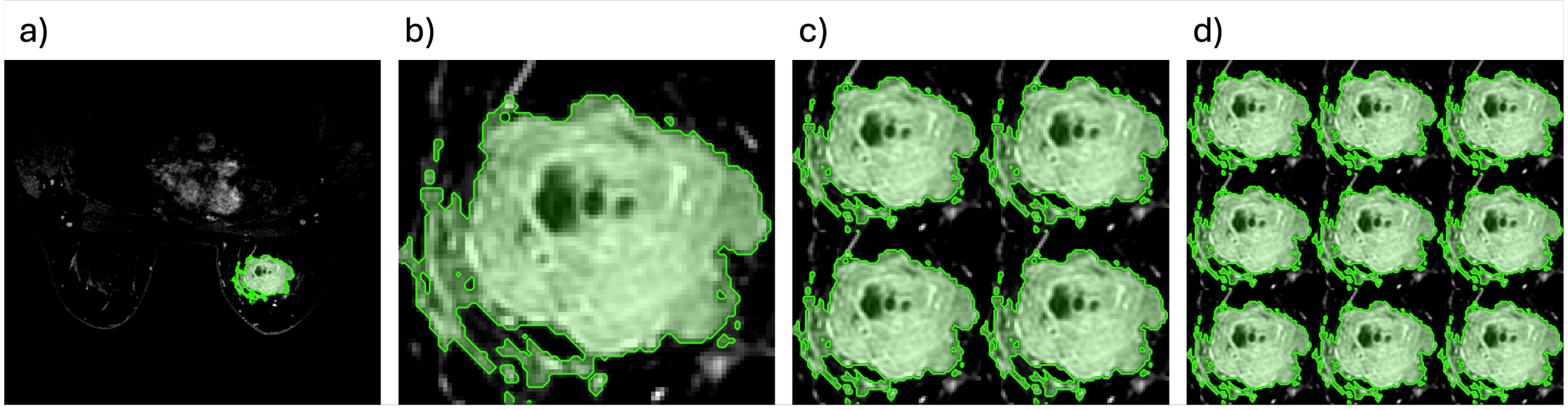

**Figure 2. Axial slices showing how the 3D computational-performance inputs were constructed. (a) Source image with the VOI shown in green. (b) Baseline crop. (c, d) Larger inputs created by repeating the crop two and three times along each spatial axis. No interpolation was used, and the repeated inputs do not represent enlarged lesions.**

## Computational environment

All computational-performance tests ran on one x86_64 node with two Advanced Micro Devices (AMD) EPYC 7742 processors. Each processor had 64 physical cores and two threads per core, giving 256 logical CPUs. The reported processor frequency was 1.5–2.25 GHz, with frequency boost enabled. CUDA tests used an NVIDIA A100 GPU with 40 GB of device memory. Direct CPU-CUDA comparisons used the same node.

# Results

## IBSI compliance and comparison with other tools

The IBSI-1 evaluation included 825 tests, covering 165 high-consensus features under five IBSI-specified configurations. The IBSI-2 source listed 396 possible tests; 73 without numerical reference values were excluded, leaving 323 tests under 22 filter configurations.

The Flash-Radiomics CPU and CUDA backends each passed all 825 IBSI-1 and 323 IBSI-2 tests under the corresponding IBSI comparison procedure. Each backend therefore passed 1,148 of 1,148 reference tests. For IBSI-1, 710 comparisons used values returned directly by Flash-Radiomics. The other 115 used quantities derived from direct shape results and configuration-specific image or mask information, as described in Methods. MIRP also passed all 825 IBSI-1 and 323 IBSI-2 tests.

PyRadiomics produced comparable results for 510 of the 825 IBSI-1 tests, passing 483 and failing 27. The other 315 tests were unsupported. For IBSI-2, PyRadiomics supported 30 of the 323 tests in the two unfiltered configurations, passing 28 and failing 2. Of the 293 unsupported tests, six involved three unavailable statistical features, and 287 required filters that were unavailable or not directly comparable. All 15 feature names used in the supported comparisons were matched. Figure 3 therefore shows both reference-test outcomes and the number of supported features and filters.

Seven PyRadiomics failures were related to the definition of kurtosis. Five IBSI-1 and two IBSI-2 values were approximately 3 units above their reference values. Subtracting 3 brought all seven within their respective tolerances, although the primary analysis retained the original values. The remaining 22 IBSI-1 failures were texture features from configuration C. This 3D lung CT configuration used trilinear interpolation to 2-mm isotropic voxels, an intensity range of −1,000 to 400 HU, and fixed-bin-width discretization with a 25-HU bin width.

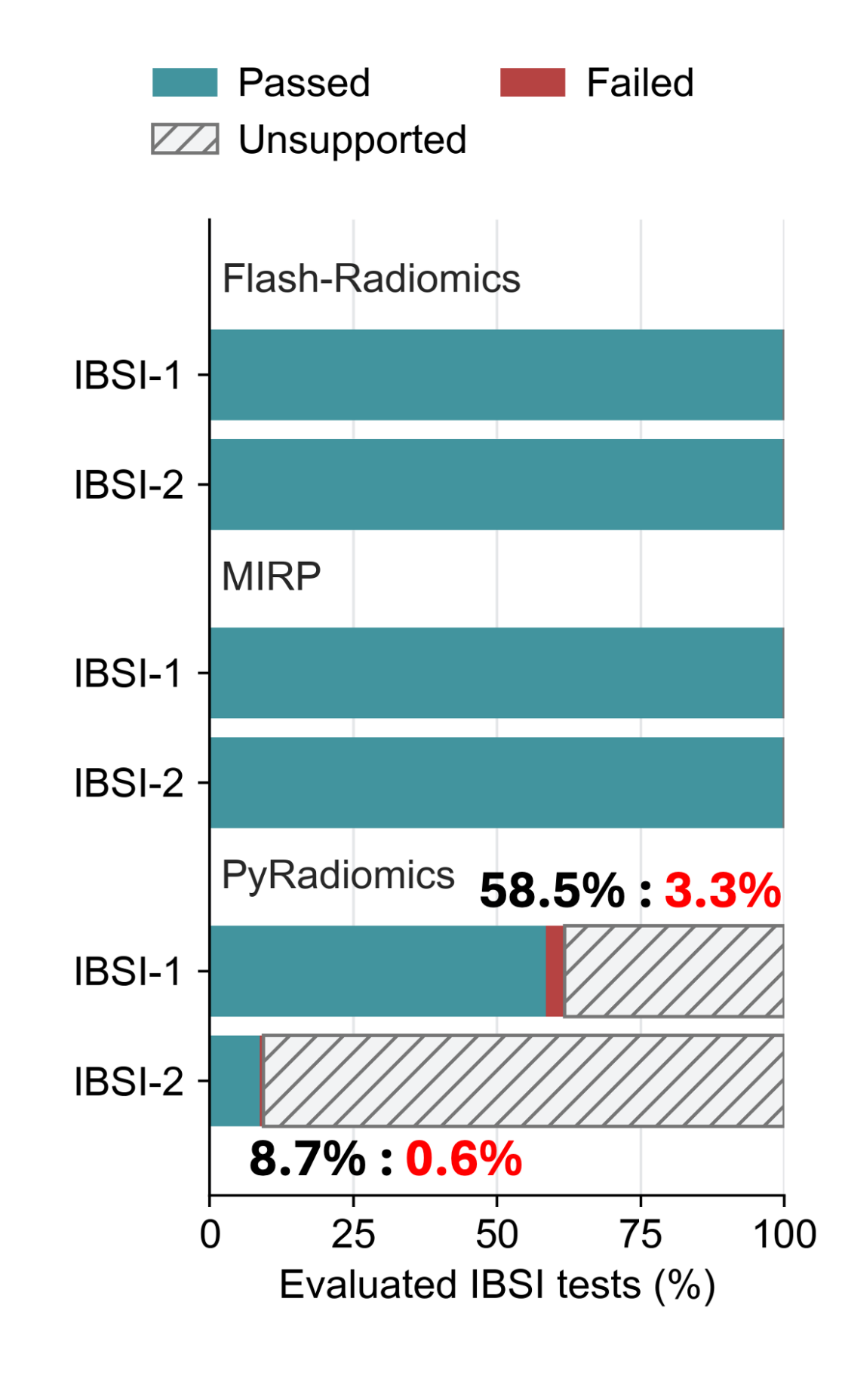

**Figure 3. Scalar IBSI reference-test outcomes for Flash-Radiomics, MIRP, and PyRadiomics. Bars show the percentages of the complete IBSI-1 and evaluated IBSI-2 reference sets that passed, failed, or were unsupported. Unsupported means that the tool did not provide a directly comparable feature or filter. Spatial maps are not included because the evaluated IBSI benchmarks do not provide reference spatial radiomic feature maps.**

## CPU-CUDA concordance

All 1,148 scalar CPU-CUDA pairs were concordant. Of these, 874 were exact and 274 were non-exact but concordant. For IBSI-1, 682 of 825 pairs were exact and 143 were non-exact but concordant. For IBSI-2, 192 of 323 pairs were exact and 131 were non-exact but concordant.

The largest relative difference was $8.89 \times 10^{-12}$, below the $1 \times 10^{-11}$ limit but above $1 \times 10^{-12}$. Both backend values for this pair passed the corresponding IBSI reference test. The largest absolute difference was 0.002453 for a large energy value; this pair passed because its relative difference was within the defined limit.

Spatial concordance was tested with the cropped image and lesion mask from the public breast cancer MRI dataset [26]. The 93 CPU-CUDA map pairs had identical geometry and contained 50,613 numerical values per map, giving 4,707,009 paired values. Every spatial map remained concordant when its finite values were evaluated with the same limits used for scalar results. Thirty-nine maps were exact, and the other 54 were non-exact but concordant. The largest absolute difference among the non-exact values was $4.37 \times 10^{-11}$, below the $1 \times 10^{-6}$ absolute limit. Figure 4 summarizes the scalar and spatial CPU-CUDA comparisons.

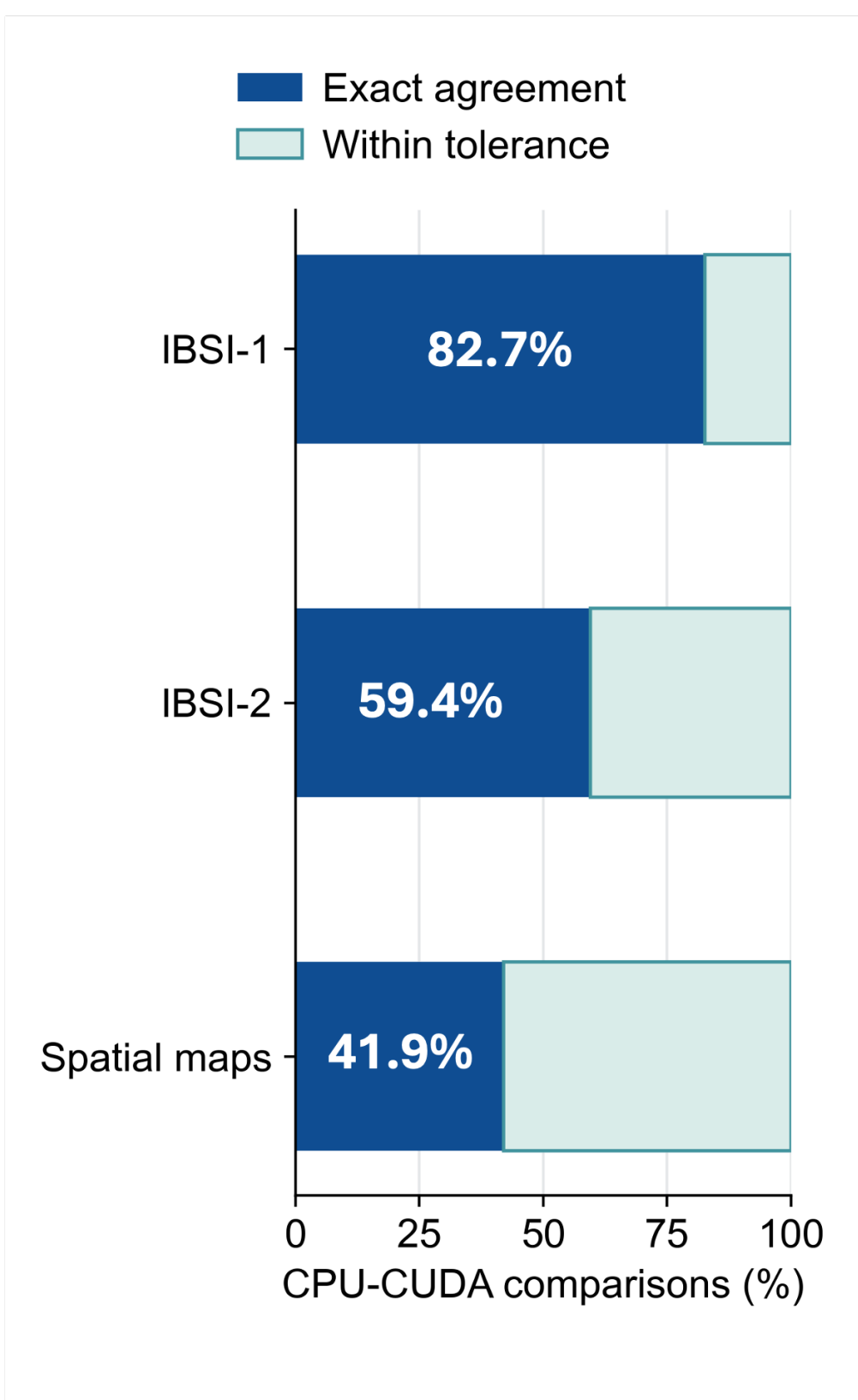

**Figure 4. CPU-CUDA concordance for scalar values and spatial maps. Bars show exact agreement and non-exact agreement within tolerance for 825 IBSI-1 scalar pairs, 323 IBSI-2 scalar pairs, and 93 spatial-map pairs. The scalar percentages classify individual value pairs. A spatial map was exact only when all finite pairs were numerically identical. It was non-exact but concordant when at least one pair differed while every finite pair met at least one numerical limit. Within tolerance denotes satisfaction of the absolute or relative limit for a scalar pair or every finite pair in a spatial map, as applicable.**

## End-to-end processing time for scalar extraction and spatial mapping

The Flash-Radiomics CPU backend was similar or faster than the hybrid CUDA backend for scalar extraction at all tested input sizes for the shared 102-feature set. The CUDA-to-CPU median-time ratio ranged from 1.000 to 1.287.

The largest input contained 10,122,600 voxels inside the VOI, corresponding to 6.18 L of repeated VOI volume within a 238.28 × 214.84 × 312.00 mm image. End-to-end processing time was 76.343 s for CPU and 81.915 s for CUDA. The fitted values of $b$ were 1.177 for CPU and 1.135 for CUDA.

All four implementations completed the scalar series. At the largest input, MIRP required 356.844 s and PyRadiomics required 14,560.914 s. Flash-Radiomics CPU was 4.7 times faster than MIRP and 190.7 times faster than PyRadiomics. Flash-Radiomics CUDA was 4.4 and 177.8 times faster, respectively. These comparisons apply only to the shared 102-feature set, preprocessing steps, tested software implementations, processing-time definition, and hardware used in this study.

The hybrid CUDA backend was faster than CPU at all five input sizes completed by both backends for the shared 93-map set. CUDA-to-CPU median-time ratios ranged from 0.313 to 0.323, corresponding to a 3.1- to 3.2-fold CUDA advantage.

At 607,356 VOI voxels, corresponding to 370.7 mL within a 95.31 × 85.94 × 117.00 mm image, CUDA required 40.193 s and CPU required 127.796 s. The largest input completed by both contained 1,214,712 VOI voxels, corresponding to 741.4 mL within a 95.31 × 128.91 × 156.00 mm image. CUDA completed this input in 79.280 s, compared with 252.791 s for CPU. CUDA therefore reduced end-to-end processing time by 68.6% and was 3.2 times faster. The fitted values of $b$ were 0.999 for CPU and 0.994 for CUDA. CUDA also completed the 10,122,600-voxel input in 663.596 s, whereas CPU reached the 5,400-s time limit.

PyRadiomics provided the same 93 named spatial maps. At 50,613 VOI voxels, corresponding to 30.9 mL, PyRadiomics required 284.562 s. Flash-Radiomics CUDA and CPU required 3.434 and 10.635 s and were therefore 82.9 and 26.8 times faster, respectively. At 101,226 VOI voxels, corresponding to 61.8 mL, PyRadiomics required 573.604 s, whereas Flash-Radiomics CUDA and CPU required 6.753 and 21.069 s. The corresponding speed differences were 84.9-fold and 27.2-fold. PyRadiomics reached the 5,400-s time limit at 404,904 VOI voxels, whereas both Flash-Radiomics backends completed this input and larger tested inputs. Figure 5 summarizes these performance results.

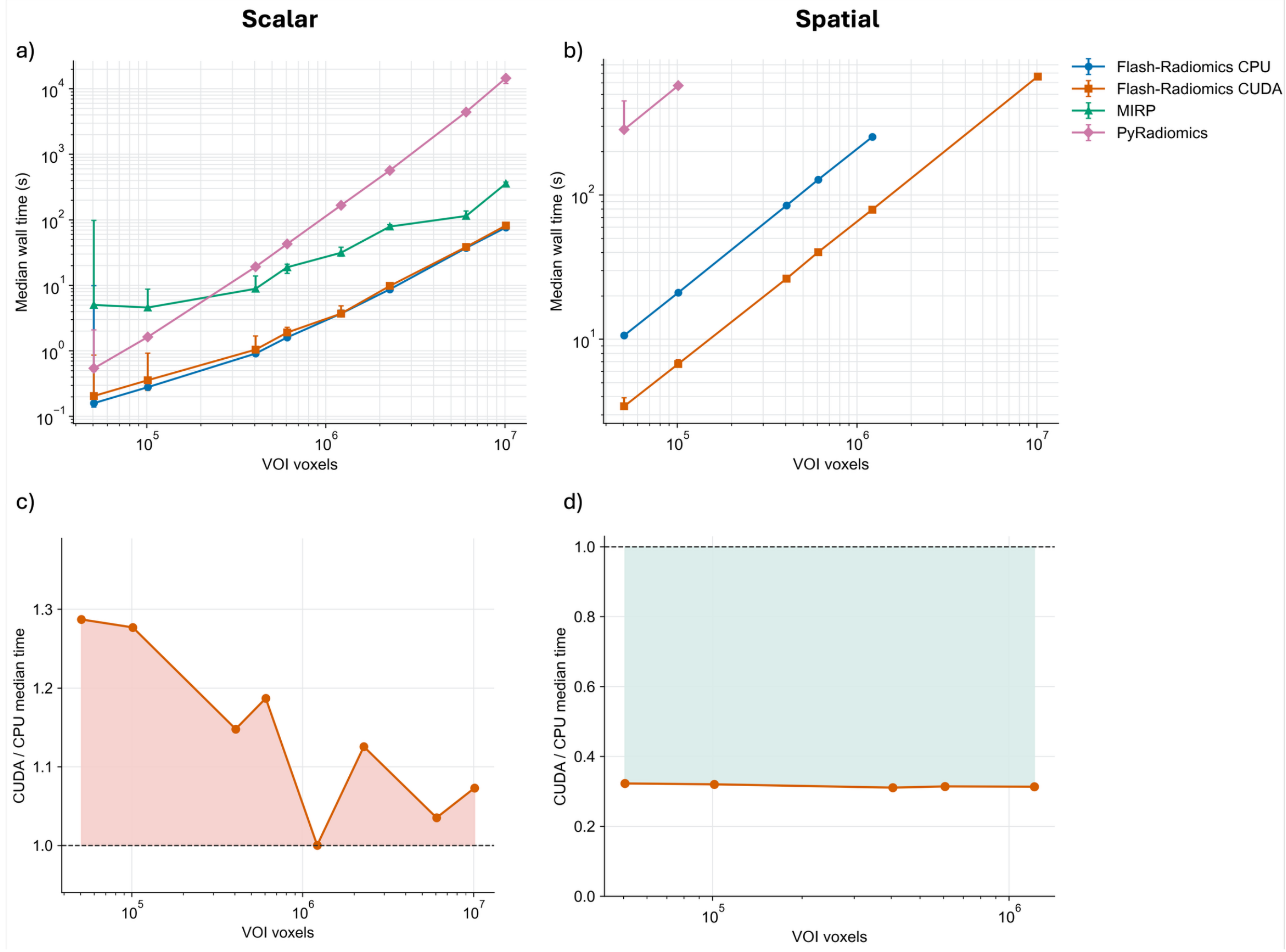

**Figure 5. End-to-end processing time as VOI size increased. (a) Median processing time for the shared 102-feature scalar set. (b) Median processing time for the shared 93-map spatial set. (c) CUDA-to-CPU median-time ratio for scalar extraction. (d) CUDA-to-CPU median-time ratio for spatial mapping. Points in panels (a) and (b) show the median of all measurements, and error bars show the minimum and maximum. Panels (c) and (d) show ratios of median times. Ratios above 1 favor CPU; ratios below 1 favor CUDA.**

## Discussion

We developed and evaluated Flash-Radiomics with two extraction modes - scalar extraction and spatial mapping, consistent feature names, and native CPU and hybrid CUDA backends. We assessed IBSI compliance, CPU-CUDA concordance, and end-to-end processing time including shared scalar and spatial feature sets using a public data set [26] and contrasted the results with publicly available radiomics tools when applicable. Our results showed that both backends passed the complete evaluated IBSI reference set, and every paired CPU-CUDA result was concordant. Backend performance depended on extraction mode: CPU was faster for scalar extraction, whereas hybrid CUDA was faster for spatial mapping. The HDF5 design retained the image, VOI mask, geometry, settings, and results in one file.

Flash-Radiomics and MIRP passed the full evaluated IBSI reference set, but PyRadiomics supported a smaller directly comparable subset. This difference highlights the need to distinguish feature coverage from numerical agreement: an unsupported feature cannot be treated as a failed calculation. Among the supported features, accounting for conventional versus excess kurtosis resolved seven PyRadiomics discrepancies, showing that differences in reported values can arise from statistical conventions rather than calculation errors. Comparisons among radiomics tools therefore require matching feature definitions and preprocessing settings before numerical agreement can be meaningfully assessed [17–20,25].

Every CPU-CUDA scalar pair and spatial map met the defined concordance limits, but exact equality was less frequent for IBSI-2 scalar pairs than for IBSI-1 scalar pairs (59.4% versus 82.7%). A possible explanation is the additional image filtering required by most IBSI-2 configurations. Because CPU and CUDA calculations can perform numerical operations in different orders, small rounding differences may arise at individual steps and carry through subsequent calculations. Adding filtering before statistical summarization provides more opportunities for these differences to appear in the final values. Exact equality was less frequent again for spatial maps (41.9%), but this percentage also reflects the requirement that all 50,613 value pairs in a spatial map match exactly. A single differing value made the entire map non-exact, however, each scalar test compared with only one value pair. The lower exact-equality percentages therefore do not establish progressively larger numerical errors as calculations become more complex. Instead, the results show that exact matches became less frequent while all differences remained within the defined concordance limits.

In the end-to-end processing time experiment, processing time increased with VOI size for both backends. CPU was similar or faster than hybrid CUDA for scalar extraction, but hybrid CUDA was approximately 3.1 to 3.2 times faster for spatial mapping at every input size completed by both backends. This change may reflect how much repeated work each extraction mode provides. Scalar extraction summarizes the VOI once for each feature, so the time saved by GPU calculations may be too small to outweigh the additional time needed to move data and coordinate CPU and GPU processing. The relative CPU advantage generally diminished as VOI size increased, which may reflect a greater computational workload relative to these additional costs. Spatial mapping repeats neighborhood calculations throughout the VOI, providing substantially more work that can be performed in parallel. The GPU can therefore offer a larger benefit relative to those additional costs. Its similar advantage across the tested spatial input sizes suggests that this benefit was already present at the smallest evaluated size. These findings support choosing the backend according to extraction mode rather than assuming that CUDA will always be faster. However, measurements of individual processing steps would be needed to confirm the reasons for the observed timing differences.

In the comparisons with other tools, Flash-Radiomics CPU was faster than MIRP and PyRadiomics for the shared scalar feature set, and both Flash-Radiomics backends were faster than PyRadiomics for spatial inputs completed by both tools. The comparisons used matched input data, preprocessing settings, result types, and shared feature sets, with the same hardware and end-to-end timing. Nevertheless, each package retained its own calculation, memory-management, and output procedures. Flash-Radiomics was designed to reduce repeated image preparation and reuse memory, which may have contributed to its shorter processing times, but the comparisons did not measure these contributions separately. The results therefore demonstrate shorter times for the end-to-end Flash-Radiomics workflows under the tested conditions without identifying a single cause or establishing a general ranking across radiomics software.

Three limitations define the scope of the findings: spatial validation, generalizability of the end-to-end computational time performance results, and clinical use. The evaluated IBSI benchmarks did not provide standardized spatial radiomic reference maps, so spatial testing relied on CPU-CUDA concordance for one image and mask from a public dataset. Agreement between the backends supports numerical consistency but cannot exclude errors shared by both implementations. The end-to-end computational time study used repeated tilings of the same crop, five measurements per input size, fixed shared feature sets, and one hardware system (Figure 2). This design allowed controlled comparisons as input size increased, but it did not capture the diversity of clinical images or computing environments. The software comparisons were also limited to tools that could produce the requested results through an automated interface. Finally, computational performance and numerical validation do not establish clinical value: the study did not evaluate associations with diagnosis, treatment response, patient outcomes, or clinical decisions. Future work should therefore prioritize independent spatial reference maps, representative multi-institutional images, additional hardware, measurements of individual processing steps, and task-specific clinical validation.

## Conclusion

Flash-Radiomics provides scalar extraction and spatial mapping with consistent feature definitions across native CPU and hybrid CUDA backends. Both backends passed all 1,148 evaluated scalar IBSI reference tests, and all 1,148 scalar pairs and 93 spatial maps were concordant between CPU and CUDA. CPU was faster for scalar extraction, whereas hybrid CUDA was faster for spatial mapping under the tested conditions; both Flash-Radiomics backends also outperformed the applicable comparator tools for the shared feature sets. HDF5 storage organizes results and associated processing information in one structured file. Overall, these findings support CPU execution for scalar radiomics and hybrid CUDA execution for computationally intensive spatial mapping.

## Code availability

The Flash-Radiomics source code is available at https://github.com/Ding3LI/flash-radiomics. The Python package is also released on the Python Package Index (PyPI).

## Data availability

The IBSI reference data and benchmark values are available from the sources cited in [17,18,24]. The breast MRI image and expert lesion segmentation were obtained from the public dataset [26]. All data used in this study is available at https://github.com/Ding3LI/flash-radiomics.

## Ethical statement

This study did not involve the recruitment of human participants or the use of identifiable patient data. This study analyzed publicly available IBSI benchmark data and breast MRI data with lesion segmentation [17,18,26]. No participants were recruited and no new patient data were collected.

## Funding

This research received no funding or grants.

## CRediT authorship contribution statement

**Shanli Ding:** Conceptualization, Data curation, Formal analysis, Methodology, Software, Validation, Visualization, Writing – original draft, Writing – review & editing. **Yiyi Hu:** Conceptualization, Formal analysis, Methodology, Visualization, Writing – original draft, Writing – review & editing. **Ziyu Fu:** Conceptualization, Formal analysis, Methodology, Validation, Writing – review & editing. **Chia-Hsin Lin:** Formal analysis, Methodology, Validation, Writing – review & editing. **Ruihan Luo:** Conceptualization, Formal analysis, Visualization, Writing – review & editing. **Jaehee Chun:** Conceptualization, Formal analysis, Software, Writing – review & editing. **Xinyue Zhang:** Data curation, Software, Validation, Writing – review & editing. **Osama Mawlawi:** Conceptualization, Methodology, Project administration, Resources, Supervision, Writing – original draft, Writing – review & editing.

## Declaration of competing interest

The authors declare no competing financial interests or personal relationships that could have influenced the work reported in this paper.

## Acknowledgements

Declaration of generative AI and AI-assisted technologies in the manuscript preparation process: During the preparation of this work, the authors used ChatGPT 5.6 in order to improve the language, clarity and readability of the manuscript. After using this tool, the authors reviewed and edited the content as needed and take full responsibility for the content of the published article.

## References

[1] P. Lambin, E. Rios-Velazquez, R. Leijenaar, et al., Radiomics: Extracting more information from medical images using advanced feature analysis, Eur. J. Cancer 48 (2012) 441–446. https://doi.org/10.1016/j.ejca.2011.11.036.

[2] H.J.W.L. Aerts, E. Rios Velazquez, R.T.H. Leijenaar, et al., Decoding tumour phenotype by noninvasive imaging using a quantitative radiomics approach, Nat. Commun. 5 (2014) 4006. https://doi.org/10.1038/ncomms5006.

[3] R.J. Gillies, P.E. Kinahan, H. Hricak, Radiomics: Images are more than pictures, they are data, Radiology 278 (2016) 563–577. https://doi.org/10.1148/radiol.2015151169.

[4] K. Bernatowicz, F. Grussu, M. Ligero, et al., Robust imaging habitat computation using voxel-wise radiomics features, Sci. Rep. 11 (2021) 20133. https://doi.org/10.1038/s41598-021-99701-2.

[5] J.J.M. van Griethuysen, A. Fedorov, C. Parmar, et al., Computational radiomics system to decode the radiographic phenotype, Cancer Res. 77 (2017) e104–e107. https://doi.org/10.1158/0008-5472.CAN-17-0339.

[6] A. Zwanenburg, S. Löck, MIRP: A Python package for standardised radiomics, J. Open Source Softw. 9 (2024) 6413. https://doi.org/10.21105/joss.06413.

[7] A.P. Apte, A. Iyer, M. Crispin-Ortuzar, et al., Technical Note: Extension of CERR for computational radiomics: A comprehensive MATLAB platform for reproducible radiomics research, Med. Phys. 45 (2018) 3713–3720. https://doi.org/10.1002/mp.13046.

[8] C. Nioche, F. Orlhac, S. Boughdad, et al., LIFEx: A freeware for radiomic feature calculation in multimodality imaging to accelerate advances in the characterization of tumor heterogeneity, Cancer Res. 78 (2018) 4786–4789. https://doi.org/10.1158/0008-5472.CAN-18-0125.

[9] LIFEx, Documentation, https://www.lifexsoft.org/index.php/resources/documentation (accessed 8 September 2026).

[10] J. Lisowski, P. Tyrakowski, S. Zyguła, K. Kaczmarski, PyRadiomics-Cuda: 3D features extraction from medical images for HPC using GPU acceleration, in: M. Paszynski, A.S. Barnard, Y.J. Zhang (Eds.), Computational Science – ICCS 2026 Workshops, ICCS 2026, Lecture Notes in Computer Science, vol. 16786, Springer, Cham, 2026, pp. 257–265. https://doi.org/10.1007/978-3-032-29912-3_20.

[11] Y. Jiao, O.M. Ijurra, L. Zhang, D. Shen, Q. Wang, cuRadiomics: A GPU-based radiomics feature extraction toolkit, in: H. Mohy-ud-Din, S. Rathore (Eds.), Radiomics and Radiogenomics in Neuro-oncology, RNO-AI 2019, Lecture Notes in Computer Science, vol. 11991, Springer, Cham, 2020, pp. 44–52. https://doi.org/10.1007/978-3-030-40124-5_5.

[12] Y. Jiao, cuRadiomics source code, GitHub, https://github.com/jiaoyining/cuRadiomics (accessed 19 August 2026).

[13] L. Rundo, A. Tangherloni, S. Galimberti, et al., HaraliCU: GPU-powered Haralick feature extraction on medical images exploiting the full dynamics of gray-scale levels, in: V. Malyshkin (Ed.), Parallel Computing Technologies, PaCT 2019, Lecture Notes in Computer Science, vol. 11657, Springer, Cham, 2019, pp. 304–318. https://doi.org/10.1007/978-3-030-25636-4_24.

[14] L. Rundo, A. Tangherloni, P. Cazzaniga, et al., A CUDA-powered method for the feature extraction and unsupervised analysis of medical images, J. Supercomput. 77 (2021) 8514–8531. https://doi.org/10.1007/s11227-020-03565-8.

[15] A. Tangherloni, L. Rundo, HaraliCU source code, GitHub, https://github.com/andreatangherloni/HaraliCU (accessed 19 August 2026).

[16] L. Zhang, D.V. Fried, X.J. Fave, et al., ibex: An open infrastructure software platform to facilitate collaborative work in radiomics, Med. Phys. 42 (2015) 1341–1353. https://doi.org/10.1118/1.4908210.

[17] A. Zwanenburg, M. Vallières, M.A. Abdalah, et al., The Image Biomarker Standardization Initiative: Standardized quantitative radiomics for high-throughput image-based phenotyping, Radiology 295 (2020) 328–338. https://doi.org/10.1148/radiol.2020191145.

[18] P. Whybra, A. Zwanenburg, V. Andrearczyk, et al., The Image Biomarker Standardization Initiative: Standardized convolutional filters for reproducible radiomics and enhanced clinical insights, Radiology 310 (2024) e231319. https://doi.org/10.1148/radiol.231319.

[19] A. Bettinelli, F. Marturano, M. Avanzo, et al., A novel benchmarking approach to assess the agreement among radiomic tools, Radiology 303 (2022) 533–541. https://doi.org/10.1148/radiol.211604.

[20] Z. Paquier, S.-L. Chao, A. Acquisto, et al., Radiomics software comparison using digital phantom and patient data: IBSI-compliance does not guarantee concordance of feature values, Biomed. Phys. Eng. Express 8 (2022) 065008. https://doi.org/10.1088/2057-1976/ac8e6f.

[21] The HDF Group, Introduction to HDF5, https://support.hdfgroup.org/documentation/hdf5/latest/_intro_h_d_f5.html (accessed 24 August 2026).

[22] PyRadiomics developers, Usage: output and voxel-based radiomics, PyRadiomics v3.0.1 documentation, https://pyradiomics.readthedocs.io/en/v3.0.1/usage.html (accessed 8 September 2026).

[23] MIRP developers, Tutorial: computing radiomics features, MIRP v2.5.0 documentation, GitHub, https://github.com/oncoray/mirp/blob/v2.5.0/docs/tutorial_compute_radiomics_features_mr.ipynb (accessed 8 September 2026).

[24] Image Biomarker Standardisation Initiative, IBSI-1 submission table and reference values, https://ibsi.radiomics.hevs.ch/assets/IBSI-1-submission-table.xlsx (accessed 8 September 2026).

[25] PyRadiomics developers, Frequently asked questions: Does PyRadiomics adhere to IBSI definitions of features?, PyRadiomics v3.0.1 documentation, https://pyradiomics.readthedocs.io/en/v3.0.1/faq.html#does-pyradiomics-adhere-to-ibsi-definitions-of-features (accessed 8 September 2026).

[26] L. Garrucho, K. Kushibar, C.-A. Reidel, et al., A large-scale multicenter breast cancer DCE-MRI benchmark dataset with expert segmentations, Sci. Data 12 (2025) 453. https://doi.org/10.1038/s41597-025-04707-4.